\documentclass[aps,prd,superscriptaddress,amsmath,amssymb,twocolumn,fleqn,nofootinbib, preprintnumbers]{revtex4-2} 

\usepackage{xcolor}
\usepackage{graphicx}
\usepackage{colordvi}
\usepackage{mathtools}
\usepackage{empheq}
\usepackage{bm}%

\usepackage{floatrow}
\usepackage{braket}
\usepackage{soul}

\usepackage{amssymb}
\usepackage{amsmath}

\usepackage{hyperref}
\def\be{\begin{equation}}
	\def\ee{\end{equation}}
\def\ba{\begin{eqnarray}}
	\def\ea{\end{eqnarray}}

\newcommand{\dd}{\mathrm{d}}

\def\popt{\texttt{PopIII} }
\def\qt{\texttt{Q3delays} }
\def\qtn{\texttt{Q3nodelays} }

\begin{document}

 \preprint{CERN-TH-2024-073}
 
	\title{The impact of a primordial gravitational wave background on LISA resolvable sources}
	
	\author{Matteo Braglia}\email{mb9289@nyu.edu}
	\affiliation{Center for Cosmology and Particle Physics, New York University, 726 Broadway, New York, NY 10003, USA}

	\author{Mauro Pieroni}\email{mauro.pieroni@cern.ch}
	\affiliation{Theoretical Physics Department, CERN, 1211 Geneva 23, Switzerland}
 
	\author{Sylvain Marsat}\email{sylvain.marsat@l2it.in2p3.fr}
	\affiliation{Laboratoire des 2 Infinis - Toulouse (L2IT-IN2P3),
Université de Toulouse, CNRS, UPS, F-31062 Toulouse Cedex 9, France}
	
	\date{\today}
	\begin{abstract}
	
The existence of a primordial stochastic gravitational wave background (SGWB) is a common prediction in various models of the early Universe. Despite constraints at different frequency ranges and claims of detection in the nHz range by Pulsar Timing Arrays, the amplitude and spectral dependence of the SGWB in the mHz range remain largely unknown. Plausible models of early Universe Physics predict a wide range of SGWB amplitudes, from undetectable to exceeding the constraints from Big Bang Nucleosynthesis. 
This paper explores the potential impact of a prominent primordial SGWB on LISA's main scientific targets. Our main analyses focuses on Massive Black Hole Binaries (MBHBs). By employing publicly available MBHB population models and state-of-the-art LISA's forecasting pipeline, we analyze the effects of the SGWB on MBHB detections. We find that the decrease in the signal-to-noise ratio induced by a strong primordial GWB can significantly reduce the number of detectable events, compromise the precision of constraints, and even hinder sky localization for some events. We also examine the impact of the SGWB on Stellar Origin Black Hole Binaries (SOBHBs) and Galactic Binaries (GBs), which are fainter sources than MBHBs. Depending on the spectral properties of the SGWB, we conclude that these sources could be either marginally affected or rendered completely undetectable.
 This largely unexplored aspect raises critical questions about the potential challenges posed by a prominent SGWB to LISA's astrophysical objectives, including MBHBs, SOBHBs, and GBs.
	\end{abstract}
	
	\pacs{Valid PACS appear here}
	\maketitle

\section{Introduction}
The Laser Interferometer Space Antenna (LISA)~\cite{LISA:2017pwj} represents a pioneering effort in the realm of gravitational wave (GW) astronomy, slated for launch in the mid-2030s. Designed to operate within the mHz frequency band, LISA's primary mission is to detect and study the merging of Massive Black Hole Binaries (MBHBs) in the distant Universe. However, its reach extends beyond this focus, encompassing a diverse array of GW sources including stellar remnant or compact object binary systems,  Double White Dwarfs (DWDs) within our galaxy and the inspiral stage of Stellar Origin Black Hole Binaries (SOBHBs), ultimately merging into the sensitivity band of current and future ground-based detectors. See, e.g.,~\cite{LISA:2022yao}, for a review of the astrophysical sources LISA will probe.

The significance of these detections cannot be overstated. They promise profound insights into fundamental questions about our Universe, enabling tests of gravity in uncharted regimes~\cite{LISA:2022kgy, Barack:2006pq, Barausse:2020rsu, Berti:2015itd}, exploration of the environments surrounding black holes~\cite{Bogdanovic:2021aav}, investigations into galaxy evolution~\cite{Breivik:2017jip, Lamberts:2018cge, Nelemans:2001hp, Ruiter:2007xx}, and even larger-scale inquiries into the expansion of our Universe~\cite{Caprini:2016qxs, Cai:2017yww, Corman:2021avn, Corman:2020pyr, LISACosmologyWorkingGroup:2019mwx}. Drawing upon the expertise of researchers from diverse fields within astrophysics and cosmology, these ambitious scientific endeavors hinge upon the accurate detection and parameter estimation of gravitational wave signals.

Current assessments of LISA's ability to detect these sources rely on noise curves derived from various factors, including the constellation's arm length, non-inertial perturbations of geodesic references, and interferometric measurement noise. As documented in several definition study reports~\cite{Seoane:2021kkk,Colpi:2024xhw}, ESA, in collaboration with the LISA consortium, continuously refines its understanding of these elements, contributing to a relatively robust estimate of LISA's sensitivity. In this paper we investigate the impact on GW parameter estimation of an additional, theoretically uncontrollable noise source: the presence of a large stochastic gravitational wave background (SGWB).

Predictions for SGWBs of astrophysical origin, arising from the superposition of many subthreshold galactic~\cite{Nissanke:2012eh, Karnesis:2021tsh, Boileau:2021sni}, or extragalactic binaries~\cite{Babak:2023lro}, are reasonably well-understood and often considered in the estimation of LISA's scientific capabilities. However, the same level of understanding does not extend to SGWBs of primordial origin, potentially leading to significant theoretical uncertainty\footnote{Extreme mass ratio inspirals also produce another (astrophysical) SGWB which is very often neglected~\cite{Bonetti:2020jku,Pozzoli:2023kxy,Naoz:2023hpz}. However, even within the theoretical uncertainties, the amplitude of this SGWB is expected to be significantly smaller than that of the primordial SGWBs considered in this work.}. Primordial SGWBs can originate shortly after the Big Bang through various types of early Universe physics, including inflation, phase transitions, or cosmic strings~\cite{Bartolo:2016ami, Caprini:2019egz, Auclair:2019wcv, Caprini:2018mtu, LISACosmologyWorkingGroup:2022jok,Braglia:2024kpo,Caprini:2024hue, Blanco-Pillado:2024aca}. The amplitude of the generated SGWBs can span from undetectable levels to extremely large, and their spectral shape varies significantly depending on the underlying mechanism. Currently, the only established constraints on the SGWB amplitude in the mHz frequency band come from observations of the Cosmic Microwave Background (CMB) and Big Bang Nucleosynthesis (BBN), limiting the (integrated over frequency) amplitude of the SGWB's energy density to be $h^2\,\Omega_{\rm GW}\,<\,1.2\times10^{-6}$ at $95$\%  CL~\cite{Pagano:2015hma}. Well-motivated models of the early universe could potentially produce anything below this upper bound. However, this constraint is not highly stringent. In fact, when translated into power spectral density, this limit is nearly 2 orders of magnitude above the nominal sensitivity of LISA at certain frequencies.

Motivated by these uncertainties, as well as the recent detection of an SGWB by Pulsar Timing Array collaborations~\cite{NANOGrav:2023gor,NANOGrav:2023hde, EPTA:2023fyk,EPTA:2023sfo,EPTA:2023xxk, Reardon:2023gzh,Zic:2023gta,Reardon:2023zen, Xu:2023wog} at smaller frequencies, which indicate that SGWB may indeed play a prominent role in uncharted frequency bands of GW astronomy, for the first time, we examine quantitatively the effects of such an SGWB on LISA sources. We begin in Section~\ref{sec:method} by extending the well-known match-filtered parameter estimation methodology to account for the presence of an SGWB. Subsequently, we conduct quantitative forecasts on parameter estimation for the most relevant sources in the LISA band in the presence of an SGWB, comparing them to the ideal, and well studied, scenario of an absent SGWB. To accomplish this, we utilize the state-of-the-art forecasting pipeline for LISA, as implemented in the \texttt{lisabeta} code. We provide an in-depth analysis of the science case for MBHBs in Section~\ref{sec:MBHB}, presenting results for several mock populations simulated based on the state-of-the-art catalogs of MBHBs. Our findings suggest that parameter estimation could be significantly impacted, leading to fewer detected sources if the SGWB is sufficiently loud. In Sections~\ref{sec:SOBH} and \ref{sec:gal}, we draw more generalized conclusions about SOBHB and Galactic Binaries (GBs), demonstrating a degradation in scientific capabilities in these scenarios as well, potentially even worse than for MBHBs.

 This work focuses on the most extreme scenario where the SGWB saturates the CMB+BBN bound. As such our results represent the most pessimistic scenario for what concerns the detection of individual GW signals with LISA and provide an upper bound on the degradation due to the unexpected presence of an SGWB. Nonetheless, given that LISA will target a completely unexplored frequency range, we must remain open to unexpected phenomena. We believe that our findings underscore the importance of considering the potential impact of an SGWB seriously and warrant further study in this area.

\section{Methodology}
\label{sec:method}

In this section, we outline our forecast methodology, beginning with a review of widely used data analysis techniques. We then discuss the incorporation of the primordial SGWB as an additional source of noise for LISA, along with illustrations of representative primordial signals consistent with existing experimental constraints.

{\bf Data and likelihood.} Let us start by considering the case where no SGWB is present in the data. Moreover, let us work under the idealized assumption that each data stream has been cleaned of glitches and other instrumental artifacts. Denoting the time-domain data stream for each of the 3 LISA channels as a combination of signal and noise, $d_i(t) = s_i(t) + n_i(t)$, where $i$ ranges over the orthogonal combinations\footnote{Assuming the LISA configuration to be a rigid equilateral triangle, with the same noise in all links, it can be shown that the signal and noise are simultaneously diagonal in such basis, which is defined through a frequency-independent rotation starting from the Michelson XYZ basis~\cite{Adams:2010vc}. Moreover, under these assumptions, the T channel is nearly signal-insensitive at low frequencies, making it a null channel that can be used to monitor the noise.} $A$, $E$, and $T$ (see, e.g., Ref.~\cite{Prince:2002hp}), we can express the data in the frequency domain by taking a Fourier transform of the time-domain data as:
\begin{equation}
	\tilde{d}_i(f) = \int_{-T/2}^{T/2} dt \, e^{2\pi i f t} d_i(t).
\end{equation}
In the following, we assume the noise to be Gaussian and stationary with the ensemble variance given by:
\begin{equation}
	\langle \tilde{n}_i(f) \tilde{n}_j^*(f') \rangle = \frac{1}{2} \delta(f-f') \delta_{ij} S_{n,\,i}(f),
\end{equation}
where $\tilde{n}_i(f)$ denotes the noise in channel $i$ in the frequency domain, and $S_{n,\,i}(f)$ is the single-sided noise power spectral density (PSD) for TDI channel $i$. Details regarding noise PSDs for each channel can be found, e.g., in Ref.~\cite{Babak:2021mhe}. 

While the noise is a stochastic variable with zero mean, which can only be characterized through its variance, the signal $s_i(t)$ is a deterministic variable. Moreover, $s_i(t)$ is a combination of the ``true'' GW signal, typically expressed in terms of a waveform~\footnote{It is important to note that our template waveform, $h(t, \bm{\theta})$, serves as an approximation of the physical signal. In the subsequent analysis, we make the simplifying assumption that our template waveforms perfectly match the astrophysical waveforms. See e.g.~\cite{Pitte:2023ltw,Speri:2022upm} for studies on this topic.} $h(t, \bm{\theta}_{\rm GW})$ depending on some physical parameters $\bm{\theta}_{\rm GW}$ characterizing the system, modulated onto each of the data streams by the detector response function $r_i(t, \bm{\theta}_{\rm r})$ with $\bm{\theta}_{\rm r}$ collectively denoting the sky-localization and the orientation of the source (for details on the computation of the LISA response function, see, e.g., Ref~\cite{Smith:2016jqs, Smith:2019wny, Babak:2021mhe, Flauger:2020qyi}). In the following, we will collectively denote the signal parameters with $\bm{\theta}$, and the signal with $s_i(\bm{\theta})$. In our analysis, we employ IMRPhenomXHM as a waveform approximant~\cite{Garcia-Quiros:2020qpx,Garcia-Quiros:2020qlt}.

An optimal estimator for the signal can be defined by taking advantage of the difference in the statistical properties of the signal and noise. Typically, this is done using the standard matched-filter inner product~\cite{Flanagan:2005yc}:
\begin{equation}
	(a|b) = 2 \int_{f_1}^{f_2} \frac{a(f)b^*(f)+a^*(f)b(f)}{S_n (f)} \, \textrm{d}f ,
\end{equation}
through which the log-likelihood can be expressed as:
\begin{equation}
\label{eq:likelihood}
	\ln \mathcal{L} = - \sum_{i=A,\,E,\,T} \frac{1}{2} \left( d_i - s_i(\bm{\theta})  | d_i - s_i(\bm{\theta})  \right),
\end{equation}
i.e., as a Gaussian log-likelihood for the noise, where three matched-filter inner products are evaluated with the PSD of the three TDI channels.

{\bf Parameter estimation: posterior sampling and Fisher formalism.} We employ two methodologies to forecast constraints on the parameters $\bm{\theta}$ starting from the likelihood above.

The first approach involves calculating the posterior distribution of the model parameters by directly sampling the likelihood. We obtain the full posterior distribution by multiplying the likelihood by the prior on the parameters using Bayes theorem $p\left( \bm{\theta} | d \right) \propto p(d | \bm{\theta})p(\bm{\theta})$, where we have dropped the evidence, as we will not be concerned with issues related to model selection. We then estimate errors on each parameter by marginalizing over the others, as customary. In practice, we calculate the likelihood utilizing the \texttt{lisabeta} code introduced in Ref.~\cite{Marsat:2020rtl}. To produce samples from the posterior, we use \texttt{ptemcee}~\cite{Vousden:2016eeu}, an ensemble sampler~\cite{Goodman-Weare:2010,Foreman-Mackey:2012any} with adaptive parallel tempering.

While sampling the posterior provides a fully Bayesian procedure to calculate any posterior distribution, regardless of its structure, it is in practice computationally consuming, and it would become quickly too demanding to scale the forecast to, e.g., a whole population of MBHBs. For this reason, in order to quickly estimate the uncertainty in the parameter reconstruction, we also pursue an alternative route and employ a simpler Fisher formalism. Defining the Fisher Information Matrix (FIM) as 
\begin{equation}
	F_{lk} \equiv   \left(  \frac{ \partial \tilde{h}^{\rm th} (f, \bm{\theta}) }{ \partial \theta_l }  \Biggr\rvert \frac{ \partial \tilde{h}^{\rm th} (f, \bm{\theta}) }{ \partial \theta_l } \right)   \Biggr\rvert_{\bm{\theta} = \bm{\theta}_0 } \; .
\end{equation}
we can approximate the likelihood in the vicinity of its  maximum $\bm{\theta}_0$ as
\begin{equation}
\ln \mathcal{L}\simeq -\frac{1}{2} F_{lk}\, \Delta \theta_l \Delta \theta_k,
\end{equation}
where $\Delta \theta_i$ is the deviation of a parameter from its true value. We can then invert the FIM to compute the covariance matrix $\Sigma_{kl}=F_{ij}^{-1}$. As customary, the confidence intervals on the model parameters $\bm{\theta}$ can be drawn from the covariance matrix $\Sigma_{lk}$. In particular, we estimate the (marginalized) error on each parameter using the Kramer-Rao bound $\sigma(\theta_i)\geq\sqrt{\Sigma_{ii}}$.

Let us emphasize that Fisher forecasts rely on the assumption of a Gaussian likelihood in the parameters $\bm{\theta}$, which is not always the case, especially when degeneracies are present. In general, this condition is more likely to be met if the signal is sufficiently loud~\cite{Vallisneri:2007ev}.

{\bf Inclusion of an SGWB.} Let us proceed by discussing the inclusion of an SGWB in the data stream and how it would affect the measurement of the individual source parameters. The key point is to realize that, for individual event characterization, an SGWB can effectively be interpreted as an additional noise source that, due to the response function, affects the different TDI channels in different ways. In practice, for what concerns the characterization of signals from individual sources, in any of the TDI channels, the instrumental noise is replaced by \footnote{See also Refs.~\cite{Reali:2022aps,Reali:2023eug} for a similar approach.}:
\begin{equation}
\tilde{n}_i\mapsto\tilde{n}^{\rm TOT}_i=\tilde{n}_i+\tilde{\sigma}_i,
\end{equation}
where $\tilde{\sigma}_i$ is the SGWB contribution in the TDI channel $i$, which is uncorrelated with $\tilde{n}_i$. Similarly to the noise, in the AET basis, the SGWB satisfies 
\begin{equation}
	\langle \tilde{\sigma}_i(f) \tilde{\sigma}_j^*(f') \rangle \equiv \frac{1}{2} \delta \left(f-f'\right) \delta_{ij} S_{{\rm SGWB, i}}\left(f\right) \; ,
\end{equation}
where $S_{{\rm SGWB, i}}$ is a combination of the SGWB PSD and the (sky average of the square of) detector response function\footnote{For details on the projection of the SGWB onto the data stream see, e.g., Refs.~\cite{LISACosmologyWorkingGroup:2022kbp, Caprini:2024hue, Blanco-Pillado:2024aca}}. Using the formula
\begin{equation}
	h^2  \Omega_{\textrm{GW}}(f) \equiv \frac{4\pi^2 f^3}{3H_0^2}  S_{{\rm SGWB}}(f )  \;, 
\end{equation}
where $H_0$ is the present value of the Hubble parameter, we can relate the PSD of the SGWB to its energy density, which is often easier to calculate from a given theoretical model. With these results, it is easy to extend our likelihood to include the effect of an SGWB by simply replacing
\begin{equation}
\label{eq:noise_SGWB_PSD}
S_{n,\,i}\mapsto S_{n,\,i}+ S_{{\rm SGWB}}
\end{equation}
in the likelihood~\eqref{eq:likelihood}. In the following, we will refer to ``No SGWB" as the case where $S_{{\rm SGWB}}$ only contains the contribution from the galactic astrophysical SGWB, sometimes also referred to as  foreground, modeled according to Ref.~\cite{Karnesis:2021tsh}.

\begin{figure}
	\centering
 \includegraphics[width=0.95\columnwidth]{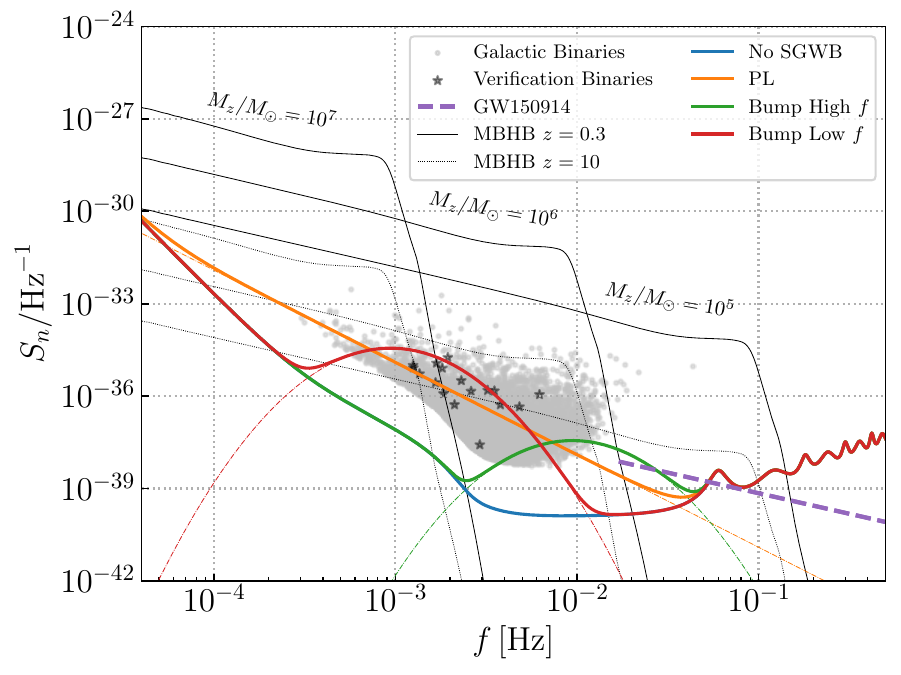}
	\caption{Power spectral density $S_n$ with and without a SGWB. For comparison, we also plot illustrative signals including MBHBs for different total masses and redshifts, a GW150914-like event~\cite{LIGOScientific:2016aoc} and galactic and verification binaries. The latter are taken from Ref.~\cite{Kupfer:2023nqx,gitlab_verification_binaries}, and we only plot those with SNR$>5$.  For MBHBs and GW150914, we only plot the + polarization of GWs for definiteness. The parameters for the SGWBs are reported in the main text and are chosen to saturate the CMB+BBN bound. All the noise curves also include galactic foregrounds. }
	\label{fig:bgs}
\end{figure}

{\bf SGWB models.} We close this section by presenting the modeling of the cosmological SGWBs that we use in our forecast.
The only current experimental limits on the amplitude of the SGWB in the mHz range are set by BBN and CMB data, which constrain the energy density of GWs at early times, through their contribution to the number of extra relativistic degrees of freedom $\Delta N_{\rm eff}$. The most updated limit is~\cite{Pagano:2015hma}:
\begin{equation}
	\label{eq:BBN_limit}
	h^2\,\int_{f_{\rm min}}^{f_{\rm max}}\,\dd\ln f\, \Omega_{\rm GW}(f)\,<\,1.2\times10^{-6}\,\,\,\,{\rm at}\,\,\,\,95\% {\rm CL},
\end{equation}
where  we take $f_{\rm min}=10^{-5}\,{\rm Hz}$ and  $f_{\rm max}=0.5\,{\rm Hz}$. For definiteness, we choose our fiducials to saturate the BBN+CMB limit~\eqref{eq:BBN_limit}. As such, the outcome of our analysis in the following Sections has to be interpreted as the most pessimistic scenario for detecting deterministic GW signals with LISA.

In this paper, we remain agnostic to the specific kind of production mechanism generating the SGWB, and we consider the following phenomenological templates:
\begin{eqnarray}
	\label{eq:PL}
	\Omega_{\rm GW}^{\rm PL} & = &A_{\rm PL}\left(\frac{f}{f_*}\right)^\alpha \; , \\
	\label{eq:Bump}
	\Omega_{\rm GW}^{\rm Bump}&=&\frac{A_{\rm Bump}}{\sqrt{2\pi}\Delta}\exp\left[-\frac{1}{2\Delta^2}\log^2 \frac{f}{f_*}\right] \; .
\end{eqnarray}
The Power Law template~\eqref{eq:PL} illustrates the effect of a broad SGWB, degrading the sensitivity of LISA over a large range of frequencies. On the other hand, the Bump template~\eqref{eq:Bump} is a proxy for any SGWB with sufficiently limited frequency support, and affects the sensitivity in a narrower range of frequencies. 

Specifically, we will consider the following benchmarks, whose impact on the mission's sensitivity  is shown in Fig.~\ref{fig:bgs}, where we also plot the strain of some typical events that are targeted by LISA: 
\begin{itemize}
	\item {\bf PL}. $A_{\rm PL}=1.27\times10^{-7},\,\alpha=0$. This case illustrates a background spanning several decades in frequency within the LISA band. Note that while the energy density is flat $\Omega_{\rm GW}$ for $\alpha=0$, its power spectral density scales as $f^{-3}$. We expect it to impact the detection and parameter estimation of MBHB spanning a large range of masses, as well as the detection of GBs and SOBHBs.
	\item {\bf Bump Low $f$}. $A_{\rm Bump}=1.27\times10^{-7},\,\Delta=0.5,\,f_*=2.2\times10^{-3}$ Hz. We choose a small width $\Delta$ in order for the SGWB to be reasonably peaked. Our choice of $f_*$ roughly corresponds to the ISCO frequency of binaries with $M_z\sim10^{7}\,M_\odot$.   As can be seen from Fig.~\ref{fig:bgs}, the SGWB peaks around the frequencies where GBs are expected to be detected and has an amplitude comparable to or even larger than their signals. However, since the SGWB signal dies off at higher frequencies, we do not expect it to have a significant effect on SOBHB detections.  
	\item {\bf Bump High $f$}. $A_{\rm Bump}=1.27\times10^{-7},\,\Delta=0.5,\,f_*=2.2\times10^{-2}$ Hz. Same as the previous benchmark, but now the choice of $f_*$ roughly corresponds to the ISCO frequency of binaries with $M_z\sim10^{6}\,M_\odot$. In addition to that, and unlike Bump Low $f$, this signal is expected to impact measurements of SOBHs, leaving almost unaltered detections of GBs.
\end{itemize}

\begin{figure*}
	\centering
	\includegraphics[width=\columnwidth]{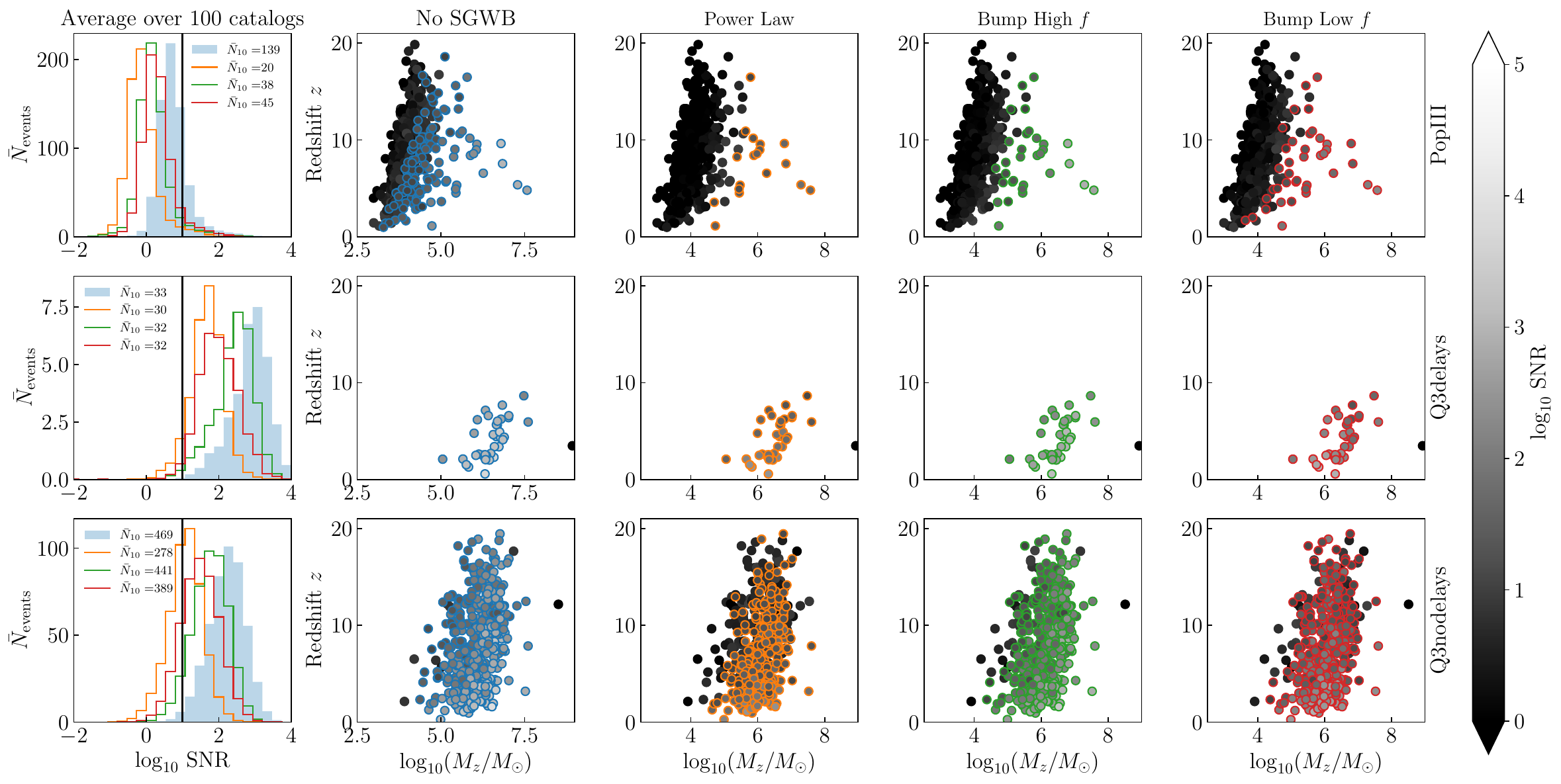}
	\caption{[First column.] We plot the ${\rm SNR}$ distribution averaged over 100 realizations of our catalogs of MBHBs populations. We emphasize the point ${\rm SNR}=10$ with a vertical black line and report the number of events detected with ${\rm SNR}>10$ in the legend. [Second, third, and fourth columns.] For a specific realization of our populations, we plot the distribution of events as a function of redshift and the total mass of the binary. The color bar indicates the ${\rm SNR}$ of each event, and we emphasize events with ${\rm SNR}>10$ with a colored circle. We plot results for \texttt{PopIII},  \texttt{Q3delays},  and \qtn populations in the first, second, and third rows respectively. Blue, orange, green, and red colors represent our results without SGWB, for {\bf PL}, {\bf Bump High $f$}, and {\bf Bump Low $f$} respectively.}
	\label{fig:scatter_PL}
\end{figure*}
\begin{figure*}
	\centering
	\includegraphics[width=\columnwidth]{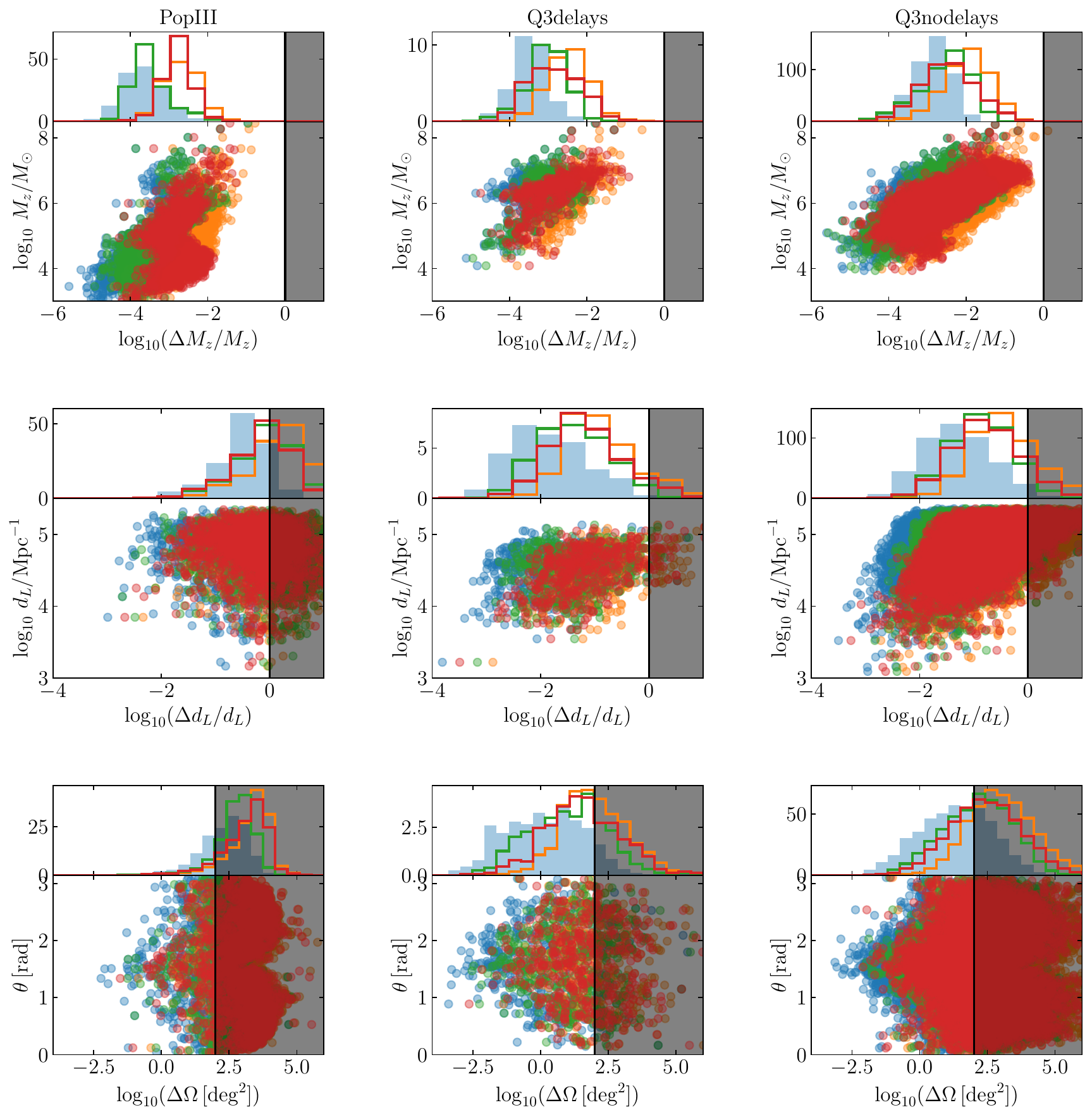}
 \includegraphics[width=\columnwidth]{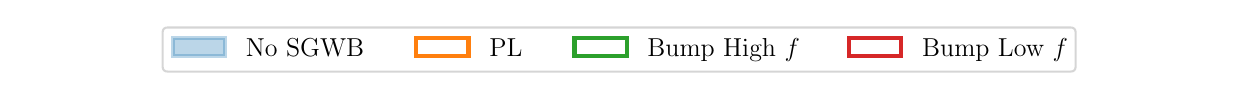}
	\caption{ Relative 68\% CL errors on the total mass $M_z$ [top] and the luminosity distance $d_L$ [center] and absolute 68\% CL errors on the sky localization $\Omega$ [bottom] from our FIM analysis. We color the region where the percentage error is $>100\%$ with black. We only perform the Fisher analysis for the events that are detected with ${\rm SNR}>10$ in the absence of an SGWB. The errors are shown as function of the fiducial values in the scatter plots, and we also provide histograms to visualize their distribution. Our results are averaged over 15 random realizations of our populations. We plot results for  \texttt{PopIII},  \texttt{Q3delays},  and \qtn populations in the first, second, and third rows respectively. Blue, orange, green, and red colors represent our results without SGWB, for {\bf PL}, {\bf Bump High $f$}, and {\bf Bump Low $f$} respectively.}
	\label{fig:Fisher}
\end{figure*}

\section{Massive Black Hole Binaries}
\label{sec:MBHB}

As highlighted in the introduction, the primary focus of our analysis is on MBHBs, aligning with one of the key science objectives of LISA. By measuring hundreds of MBHBs, LISA will contribute significantly to answering questions about the formation and evolution of these systems at high redshifts, a topic that remains a subject of ongoing investigation. In this section, our goal is to assess the impact of a significant SGWB on parameter estimation for individual MBHBs and extend this analysis to their populations.

For clarity in our subsequent discussions, we introduce the signal-to-noise ratio (SNR) of a GW event as ${\rm SNR}^2\equiv(s|s)$. Here, the scalar product is computed using the noise power spectral density (PSD) defined in Eq.~\eqref{eq:noise_SGWB_PSD}, which accommodates the inclusion of the SGWB. Additionally, we define the quantity $\bar{N}_{\rm events}$ as the number of events with an SNR greater than or equal to 10, which we establish as a threshold for detecting a GW event.

\subsection{Population analysis}

Let us begin by introducing our populations of MBHBs. We adopt the
populations presented in Ref.~\cite{Klein:2015hvg}, based on the semi-analytical model described in Ref.~\cite{Barausse:2012fy}. We note that slightly more updated predictions for MBHB populations exist (see, e.g.,~\cite{Barausse:2020kjy,Barausse:2020mdt} and references therein). However, since the data for the populations in~\cite{Klein:2015hvg} are publicly available\footnote{\href{https://people.sissa.it/~barausse/catalogs}{https://people.sissa.it/barausse/catalogs}}, we employ those for our analysis. Moreover, while results would quantitatively change for different populations, the qualitative points raised by our work would still hold. Following the methodology of Ref.~\cite{Klein:2015hvg}, we generate three distinct populations that primarily differ in the initial MBH seeds and the time delay between MBH and galaxy mergers. For each of these populations, we illustrate in Fig.~\ref{fig:scatter_PL} the distribution of the SNR of the generated events, along with their distributions as functions of redshift and (redshifted) total binary mass. To provide insight into the effects of introducing an SGWB, we plot these quantities for each SGWB benchmark discussed in the previous section.

Next, we delve into the features of these generated populations and explore the impact of an SGWB on the detectability of their sources:

\begin{itemize}
		\item {\bf \popt}: This population is based on the assumption of light MBH seeds resulting from the explosion of \popt stars and incorporates delays between MBH and galaxy mergers, making the model realistic and conservative. Notably, even if the delays were set to zero, the event rate would only change by approximately a factor of two~\cite{Klein:2015hvg}. When averaging over 100 realizations of our catalogs, we identify roughly 140 detectable events with ${\rm SNR}\geq10$. As depicted in the upper panels of Fig.~\ref{fig:scatter_PL}, this population predicts a high merger rate extending up to $z\sim20$, with many detectable events occurring at high redshifts. Arising from light seeds, most of these events involve binaries with masses ranging between $M_z\sim 10^4-10^5 M_\odot$, while only a few dozen have larger masses of around $M_z\sim 10^6-10^8 M_\odot$. The SGWB exerts a substantial impact across all three benchmarks, with the most pronounced effect observed for the PL case, where only approximately one-seventh of the events remain detectable, predominantly with larger masses. In contrast, in both the Bump Low and High $f$ scenarios, nearly double the events remain detectable, albeit with a more pronounced reduction in SNR around masses $M_z\sim10^7 M_\odot$ and $M_z\sim10^5 M_\odot$ respectively compared to the PL case, consistent with the discussion in the previous section.
	
		\item {\bf \qt}: This population is distinguished by its heavy MBH seeds originating from the collapse of protogalactic disks. Similar to \texttt{PopIII},  it accounts for the delays between MBH and galaxy mergers, lending it a realistic and conservative nature. The name "Q3" is derived from the so-called  critical Toomre parameter $Q_c$, which quantifies the stability of protogalactic disks and the likelihood that a halo hosts a black hole seed. In this model, $Q_c$ is set to 3~\cite{Klein:2015hvg}, although reasonable values typically range from 1.5 to 3 (see, e.g., Ref.~\cite{Volonteri:2007ax}). The mass and redshift distribution of events in this model differs from that of \texttt{PopIII},  with all events occurring at redshifts $z\leq10$ and involving large masses $M_z\gtrsim10^6 M_\odot$. We identified approximately 30 detectable events for this model, a number significantly lower than that for \popt. However, their SNRs are consistently high, with nearly all events remaining detectable even in the presence of a strong SGWB, although the SNR may degrade by a large extent---up to an order of magnitude in the PL case.
	
	\item {\bf \qtn}: This population mirrors \texttt{Q3delays},  but without accounting for delays between galaxy and MBHB mergers, presenting a more optimistic scenario for LISA's MBHB merger rates. The event distribution extends to high redshifts, reaching $z\sim20$, similar to \texttt{PopIII},  and involves mainly large masses $M_z\gtrsim10^6 M_\odot$, akin to \texttt{Q3delays},  but with some events ranging from $M_z\sim10^5 M_\odot$ to $M_z\sim10^8 M_\odot$. We identify nearly 500 detectable events, which is over three times the number for \popt and ten times that of \qt. Similar to \texttt{Q3delays},  $\bar{N}_{\rm events}$ does not change dramatically, although the average SNR is reduced by up to an order of magnitude.
	
\end{itemize}

The distribution of events within each population heavily influences the impact of the SGWB on the detectability of MBHBs. Populations with a significant fraction of events at masses $M_z\lesssim10^5-10^6 M_\odot$ (like \popt in our study) experience a notable reduction in detectable events. This reduction could significantly impact the reconstruction of the mass distribution of events. Conversely, populations with a predominant fraction of events at higher masses, such as \qt and \texttt{Q3nodelays}, fare better under the influence of the SGWB.

To quantify these effects more rigorously, we extend our analysis beyond simple SNR considerations to assess the degradation of parameter estimation due to the SGWB. We employ the Fisher formalism to derive constraints on three key parameters: the total mass $M_z$, the luminosity distance $d_L$, and the sky localization $\Delta\Omega$. Our results are depicted in Fig.~\ref{fig:Fisher}. These figures compare the (marginalized) error distributions for events with ${\rm SNR}\geq10$ in the absence and presence of our three benchmark SGWBs. 
Regions where the forecast relative error exceeds $100\%$ indicate non-detection. For sky localization, we use a non-detectability criterion of $\log_{10} \Delta\Omega\,[{\rm deg}^2]\geq2$. Our results are based on averaging over 15 realizations, to average over Poisson fluctuations and better represent the actual event distribution.

From Fig.~\ref{fig:Fisher}, the error distribution for the total mass $M_z$ shifts by an order of magnitude towards larger errors for all three populations. Notably, errors for \qt and \qtn populations remain relatively low compared to \texttt{PopIII},  where errors can reach up to $10\%$. 

In Fig.~\ref{fig:Fisher}, the impact of the SGWB is even more pronounced, particularly for the \popt population. We observe a degradation of the relative error by approximately one order of magnitude. While this poses no significant threat to populations dominated by high-mass events, it severely impacts populations like \popt where low-mass events are prevalent, pushing many events into non-detection territory. The scatter plot is also useful to understand the trend of the errors on the mass and confirm this argument. For \qt and \texttt{Q3nodelays}, more massive events have a worse determination of $M_z$ than low-mass ones. Indeed, as can be seen from Fig.~\ref{fig:scatter_PL}, larger-mass events in these populations are more distant than lighter ones. As for \texttt{PopIII},  many low-mass events are associated with high redshift, so that many of the scatters, especially in the Bump low $f$ and Power Law cases, are moved to the right, and the peaks of the histogram consequently shift to a larger error.

The error on the distance also varies depending on the population. Faraway sources typically have larger errors. However, while for populations such as \qt and \texttt{Q3nodelays}, the error on the distance does not seem to preclude their detectability, many of the high-redshift events in \popt have a lower mass (see also second column of Fig.~\ref{fig:scatter_PL} in the previous Section), and are therefore much weaker, making the relative error on the distance larger than 100\%. The scatter plots reveal a horizontal shift to the right for all populations, almost independent on the distance. This is consistent with the fact that the error on the distance effectively scales with the SNR of the measurement, which is reduced if an SGWB is present (see also discussion below).

Turning to the sky localization $\Delta\Omega$, whose forecast uncertainties are shown in the bottom panels of Fig.~\ref{fig:Fisher}, we find that constraints on this parameter are also significantly degraded by the presence of an SGWB. Even the \qt and \qtn populations are affected. For \texttt{Q3nodelays}, nearly half of the events with ${\rm SNR}\geq10$ are impacted for the Bump Low and High $f$ cases, and even more so for the PL case.

While our Fisher formalism provides a quantitative estimate of these impacts, its assumptions may break down when approaching the detection threshold (SNR$\lesssim10$), as indicated by the red points in Fig.~\ref{fig:Fisher}. For events with SNR$\lesssim10$ our results should therefore be considered just as rough order of magnitude estimates of the errors, and most conservative errors should be obtained through an MCMC forecast (see next Section). Despite this limitation, we decided to keep these events in Fig.~\ref{fig:Fisher}, in order to compare the same number of events in the cases with and without SGWB.

\begin{figure*}
	\begin{minipage}{.49\textwidth}
		\includegraphics[width=\columnwidth]{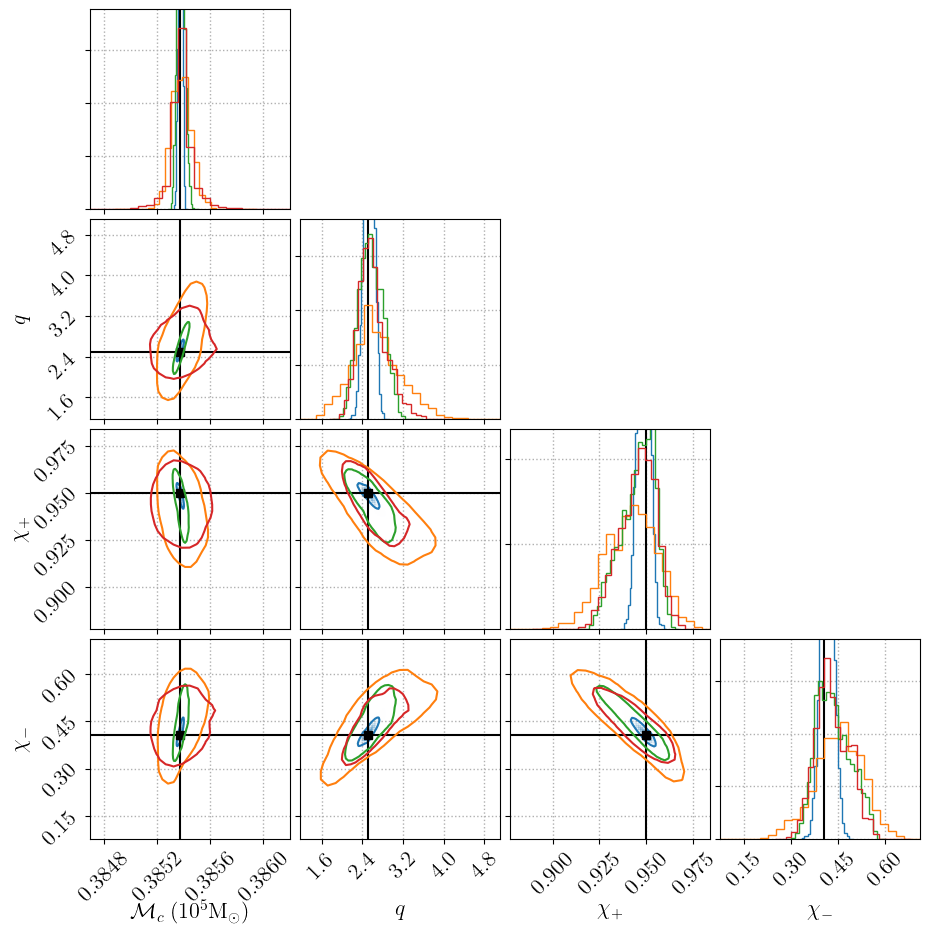}
	\end{minipage}
	\begin{minipage}{.49\textwidth}
		\includegraphics[width=\columnwidth]{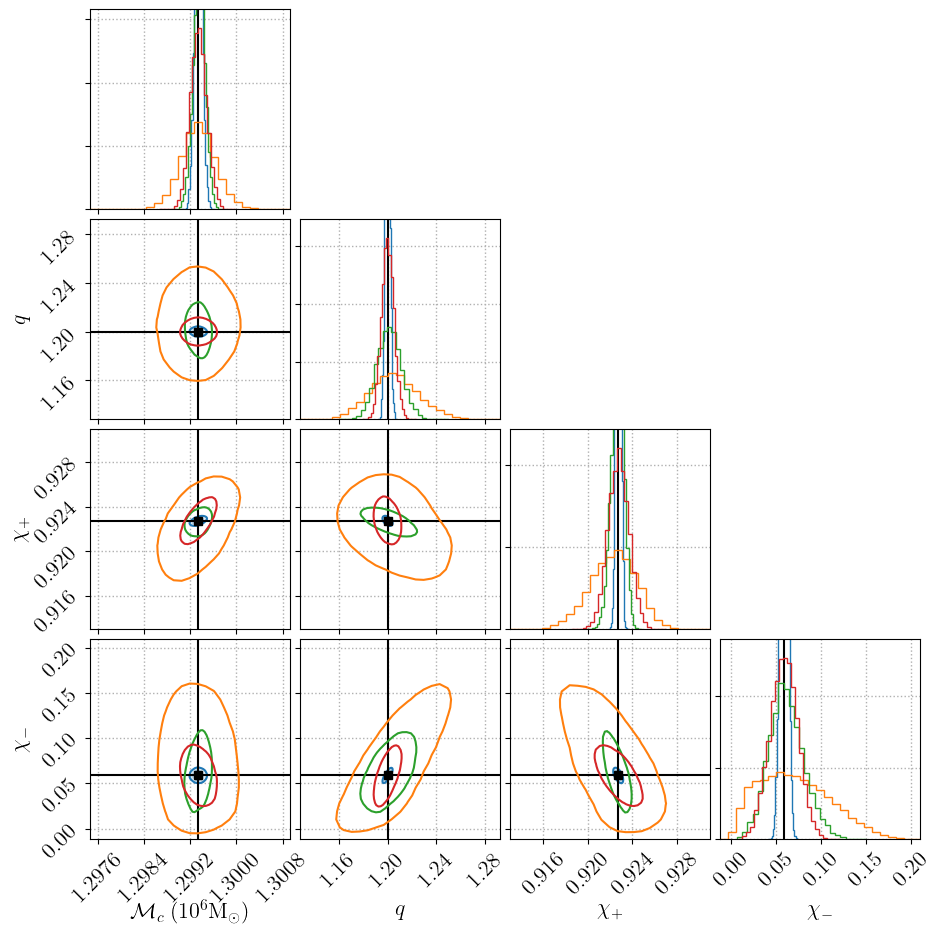}
	\end{minipage}
	\begin{minipage}{.49\textwidth}
		\includegraphics[width=\columnwidth]{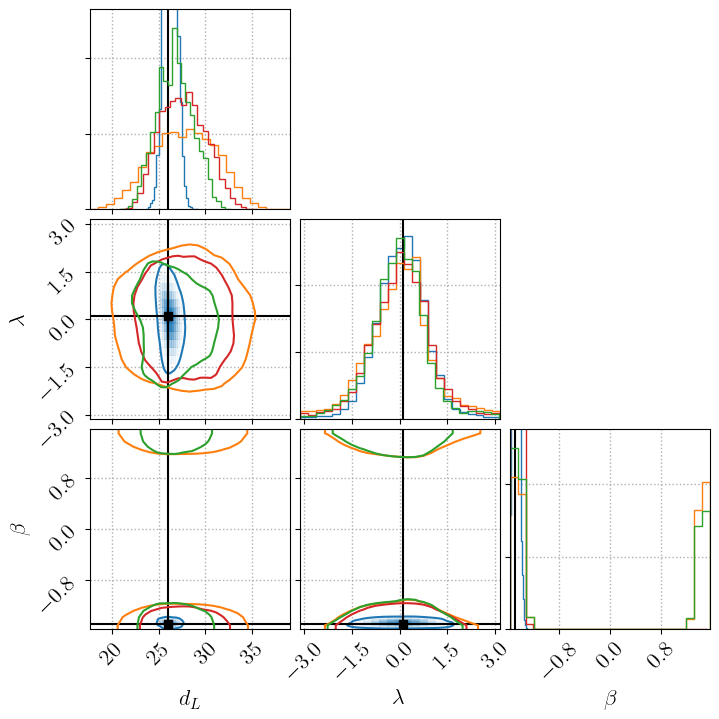}
	\end{minipage}
	\begin{minipage}{.49\textwidth}
		\includegraphics[width=\columnwidth]{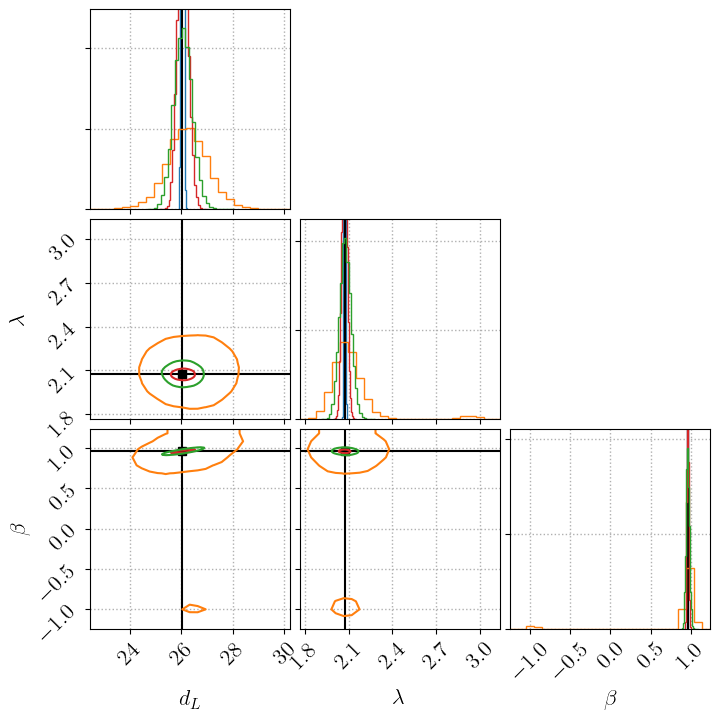}
	\end{minipage}
	\includegraphics[width=\columnwidth]{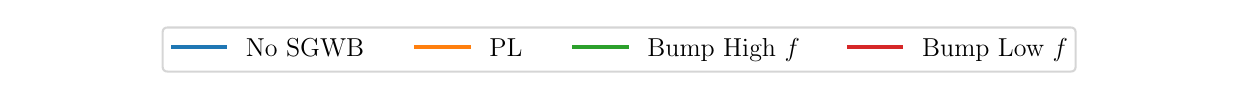}
	\caption{Results of simulated Bayesian parameter estimation for the MBHB 1 [left panels] and MBHB 2 [right panels] systems of Table~\ref{tab:MBHBparams}.  Shown is the posterior distribution on intrinsic parameters, masses, and spins [upper panels] and on the two angles for sky localization, as well as luminosity distance [lower panels]. Contours are shown at 90\% CL. Blue, orange, green, and red colors represent our results without SGWB, for {\bf PL}, {\bf Bump High $f$}, and {\bf Bump Low $f$} respectively. All intrinsic and extrinsic parameters are varied and marginalized over in the analysis. We present them in separate corner plots only for convenience.}
	\label{fig:corner}
\end{figure*}

\begin{table}
	\begin{ruledtabular}
		\begin{tabular}{c | c | c} 
			& MBHB 1  & MBHB 2  \\
			\hline
			$M_z$  & $10^5 M_\odot$ & $3\times10^6 M_\odot$  \\
			$q$  & $2.5$ & $1.2$  \\
			$\chi_1$  & $0.95$ & $0.9$  \\
			$\chi_2$  & $0.95$  & $0.95$  \\
			$z$  & $3.$ & $3.$  \\
			$t_0$  & $0.75$ &  $0.65$ \\
			$\iota$  & $2.4$ & $2.8$  \\
			$\varphi$  & $-0.48$ &  $0.$ \\
			$\lambda$  & $-1.48$ & $1.5$  \\
			$\beta$  & $-0.53$ &  $0.2$ \\
			$\psi$  & $1.8$ &  $1.66$ \\
			\hline 
			SNR no SGWB  & $228.8$ & $5096.2$  \\
			SNR power law & $29.4$ & $239.6$  \\
			SNR bump low-f  & $68.7$ & $1411.4$  \\
			SNR bump high-f  & $84.9$ & $442.1$  \\
		\end{tabular}
	\end{ruledtabular}
	\caption{\label{tab:MBHBparams} Parameters of the MBHB systems chosen for a simulated Bayesian parameter estimation.}
\end{table}

\subsection{In-depth discussion for illustrative injections}
To complement the analysis of the previous section, we delve into the detailed parameter degeneracies, employing a direct sampling approach of the likelihood distribution~\eqref{eq:likelihood} to capture potential non-Gaussian features and multimodalities of the posterior distributions. For simplicity, we focus on two representative events, whose parameters are listed in Table~\ref{tab:MBHBparams}. We select two injections to represent typical events predicted by our populations. While MBHB 1 is more likely found in populations similar to \texttt{PopIII},  it can also appear as an outlier in populations like \qt and \texttt{Q3nodelays}, and vice versa for MBHB 2. Furthermore, we choose the parameters of the events in such a way that they develop multimodalities in the distribution of the sky localization upon addition of the SGWB, a feature which is common to many events in our population and that we discuss in depth below. 

MBHB 1 and MBHB 2 are relatively close events, with $z=3$. Thus, they exemplify strong signals that LISA would detect with a quite large SNR in the absence of a large SGWB. The presence of an SGWB would still allow the detection of these two events, albeit with a significantly reduced SNR compared to the case without an SGWB. This reduction is most prominent for the PL case, for which the SNR is reduced by a factor $\sim 8$ and $\sim 21$ for MBHB 1 and MBHB 2 respectively.  We illustrate the SGWB impact on the accuracy of parameter constraints in Fig.~\ref{fig:corner}, where we present the 1D and 2D posterior distributions of some representative parameters. 

Let us start by discussing the intrinsic parameters. We focus on the spin projections along the angular momentum of the binary, $\chi_-$, $\chi_+$, on the redshifted chirp mass $\mathcal{M}_c$, and the mass ratio $q$. Visual inspection of the posterior distributions reveals Gaussian-like shapes for most parameters, with slight non-Gaussianities observed in the contours involving the spins. 
Interestingly, the position of the SGWB bump does not significantly impact the estimation of the spins for both MBHB 1 and 2. On the other hand, compared to Bump High $f$, the Bump Low $f$ scenario severely affects the estimate of the chirp mass for MBHB 1. This can be attributed to the noise level being comparable to the strain of signals like MBHB 1 during their inspiral stage, thereby reducing the ability to constrain the chirp mass of such events. Among the benchmark SGWBs considered, Bump Low $f$ provides the worst-case scenario for constraining the chirp mass of events with relatively small total masses. This situation may deteriorate further for events with masses of the order of $10^4 M_\odot$, where much of the information about the inspiral phase could be obscured by a large SGWB peaked at small frequencies. Except for this parameter, the PL case stands out as the most pessimistic, providing the least favorable constraints, particularly evident for MBHB 2, where constraints worsen by nearly an order of magnitude compared to the no SGWB scenario, diminishing the expected accuracy for such bright signals. Finally, we stress a non-trivial effect occurring in the $q-\chi_+$ plane in the MBHB 2 case. We notice that in the High $f$ Bump case, the ellipse is oriented in a different direction with respect to the PL and Low $f$ bump cases. This effect originates from the partial masking of the last part of the GW signal, which effectively modifies the relative weights of the contributions brought by the different frequencies to the total information. Since most information on the spin parameters comes from the very last part of the GW signal, the High $f$ Bump case degrades more substantially the determination of the spins compared to all other parameters.

Moving on to the discussion of extrinsic parameters, we focus on the luminosity distance to the source, $d_L$, the ecliptic latitude and longitude, denoted with $\beta_L$, and $\lambda_L$, respectively. We start by commenting on the degradation of $d_L$, which in both cases is more significant in the PL case compared to the Low and High $f$ bumps. Notice the nontrivial scaling of the degradation of the uncertainty in the determination of $d_L$ for the two bumps in the MBHB 1 and MBHB 2 cases. This behavior originates from the relative difference between the frequency of the mergers and the bumps: for MBHB 1, the two bumps only mask part of the inspiral, while for MBHB 2, the High $f$ bump masks part of the merger. The posterior distributions of $d_L$ shown in Fig.~\ref{fig:corner} and the SNR values reported in Tab.~\ref{tab:MBHBparams} are thus consistent with simple scaling arguments, which suggest that the uncertainty on $d_L$ scales inversely with the SNR.

In addition to the constraints on the distance of the events, we also notice a degradation in the sky localization of the events. 
Information about the sky localization comes from both the inspiral and merger stages of the signal. The inspiral stage contributes more significantly for low-mass systems, while the merger stage is more dominant for high-mass systems. In the presence of an SGWB affecting either low or high frequencies, lighter or heavier systems will experience a degradation in their localization information. This degradation might also give rise to multimodality in the sky localization of the source, making it challenging to precisely determine the true signal parameters even for nearby sources. This effect is manifest in the panels of the two triangle plots showing the constraints on the latitude $\beta_L$. Particularly for MBHB 1 (and also in the PL case for MBHB 2), the injected event could not be distinguished from a similar signal with a flipped sign of $\beta_L$. 

\begin{figure*}
    \centering
    \includegraphics[width=\columnwidth]{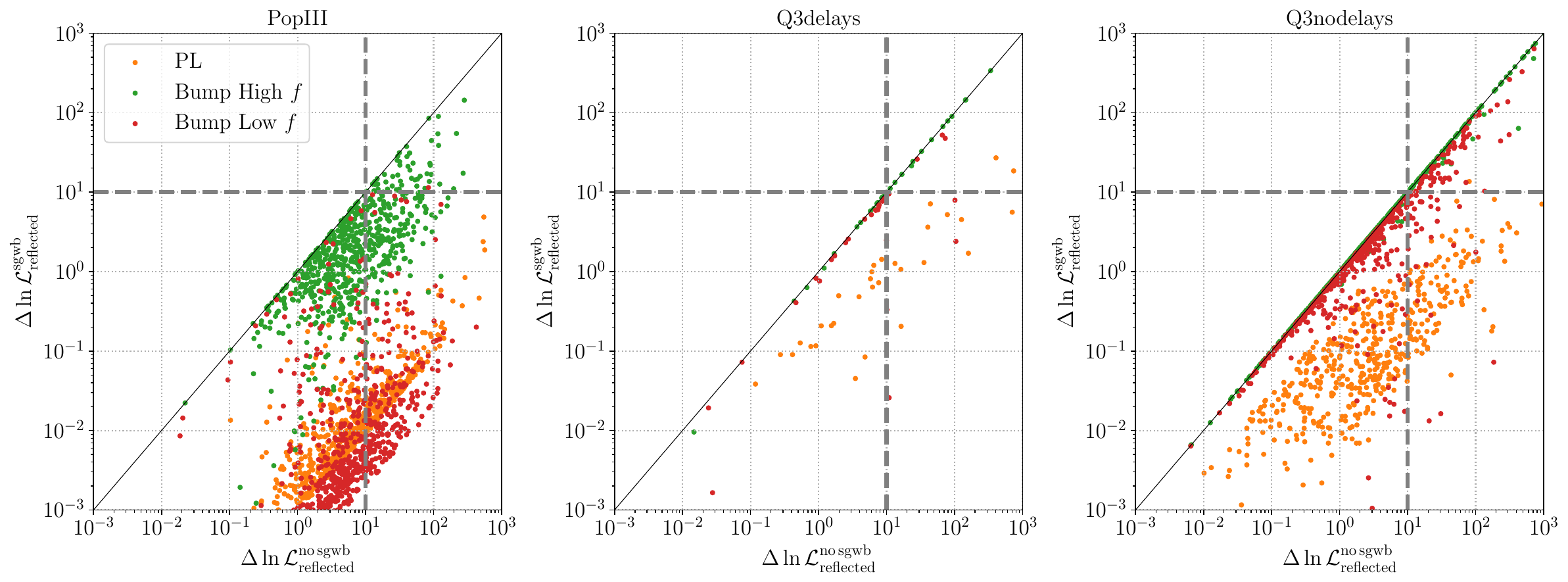}
    \caption{Scatter plot for our 3 populations showing the value of an approximate measure for sky multimodality, $\Delta \ln \mathcal{L}_{\rm reflected}$, computed in the absence of an SGWB (x-axes) versus its value computed in the presence of our 3 benchmark SGWBs (y-axes). Events on the left of the vertical dashed gray line are likely to feature multimodal sky localizations even in the absence of an SGWB. 
    Events in the bottom right quadrants, i.e. to the right of the vertical and below the horizontal dashed gray lines, are likely to feature a multimodal sky localization only when the SGWB is included. 
    }
    \label{fig:multiple_images}
\end{figure*}

The importance of this effect varies depending on the properties of the population of MBHBs. We explore its relevance for the different populations and SGWBs considered in this work in Fig.~\ref{fig:multiple_images}. To quantify this effect, we compute the difference $\Delta \ln \mathcal{L}_{\rm reflected}$ between the log-likelihood evaluated in the true parameters and the sky position most likely to develop a secondary mode, corresponding to the flipped sign of $\beta_L$. The $x$/$y$-axes of the panels in Fig.~\ref{fig:multiple_images} correspond to this quantity evaluated in the ``No SGWB'' case and in the presence of the SGWB, respectively. Although the mapping from this simplified metric to the probability support of this possible secondary mode in the sky is not simple, we can consider that for a value of $\Delta \ln \mathcal{L}_{\rm reflected}$ smaller than $\sim 10$, highlighted by a vertical/horizontal line, the secondary mode becomes compatible with the injected values, meaning it cannot be excluded by the measurement LISA will perform. We can see that, compatibly with the results shown in Fig.~\ref{fig:Fisher}, the number of sources featuring a multimodality in the sky position depends on the population/SGWB benchmark, with the High $f$ bump and \qt being consistently the less dramatic cases.

\section{Other sources in the LISA band}
Having carefully re-analyzed the science case for MBHBs, we now move on to the other main astrophysical sources in the LISA band: SOBHBs and GBs. In the next subsections, we will limit the analysis to a simple SNR computation. As we will shortly see, this would be sufficient to draw important conclusions, without employing more complex parameter estimation techniques like the one we performed for MBHBs.

\subsection{Stellar Origin Black Hole Binaries}
\label{sec:SOBH}

\begin{figure}
	\centering
	\includegraphics[width=0.95\columnwidth]{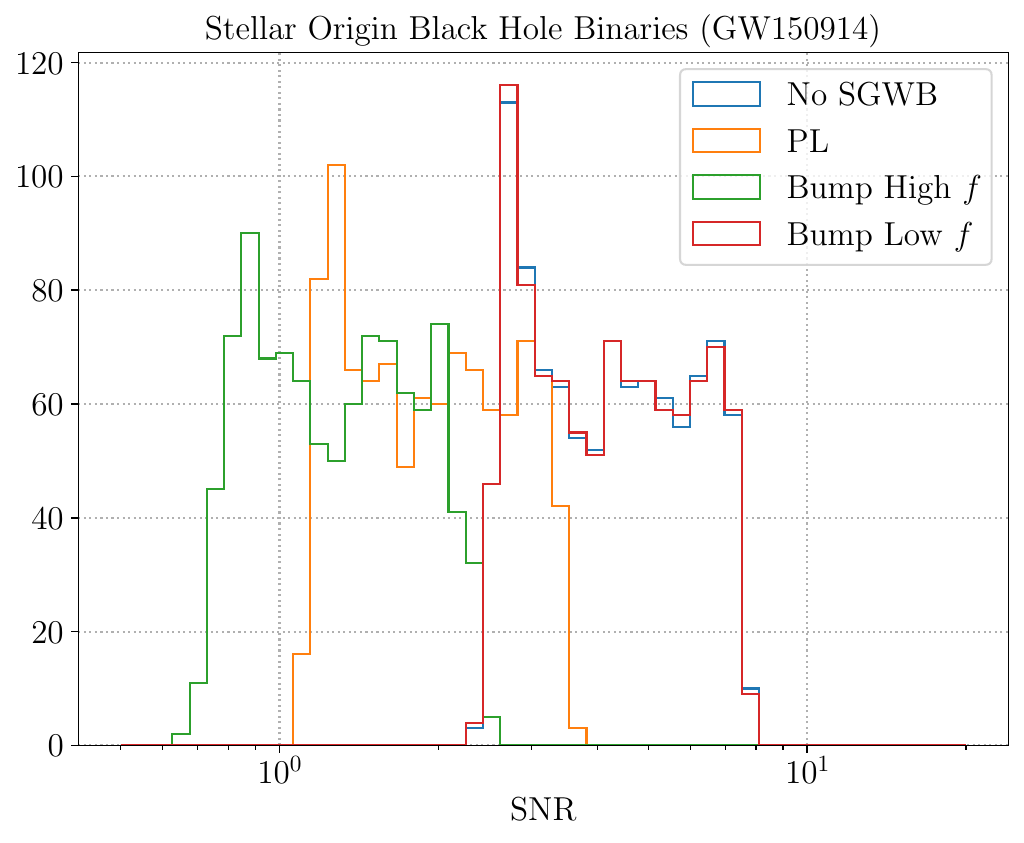}
	\caption{SNR histogram for a SOBHB system similar to GW150914, varying inclination, phase, sky position and polarization, for the different SGWB cases.}
	\label{fig:SOBBH_bgs}
\end{figure}

\begin{figure*}
	\centering
	\includegraphics[width=0.32\columnwidth]{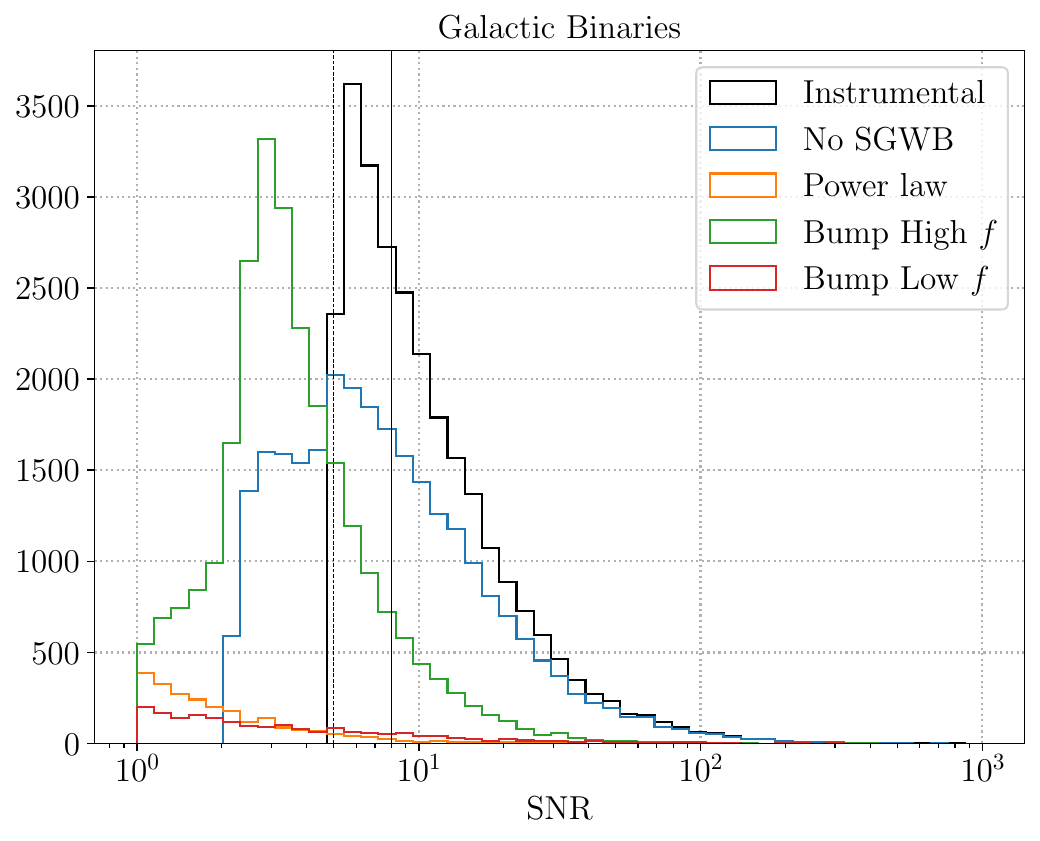}
	\includegraphics[width=0.32\columnwidth]{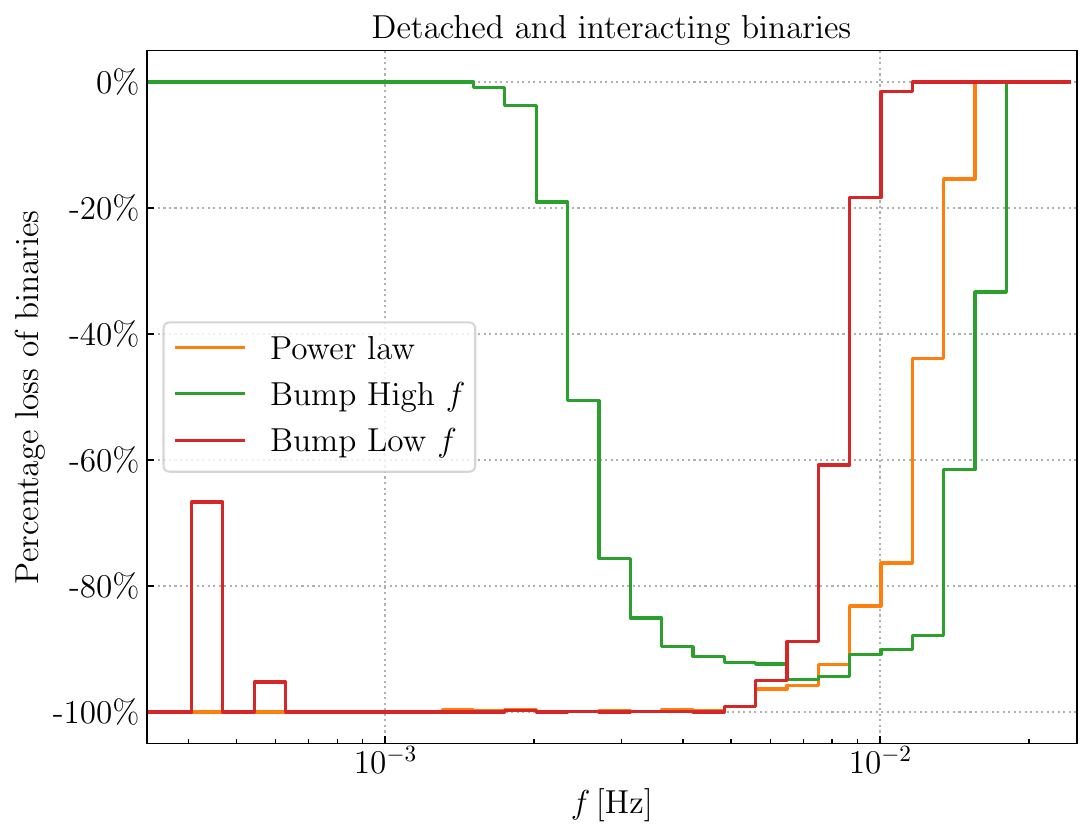}
	\includegraphics[width=0.32\columnwidth]{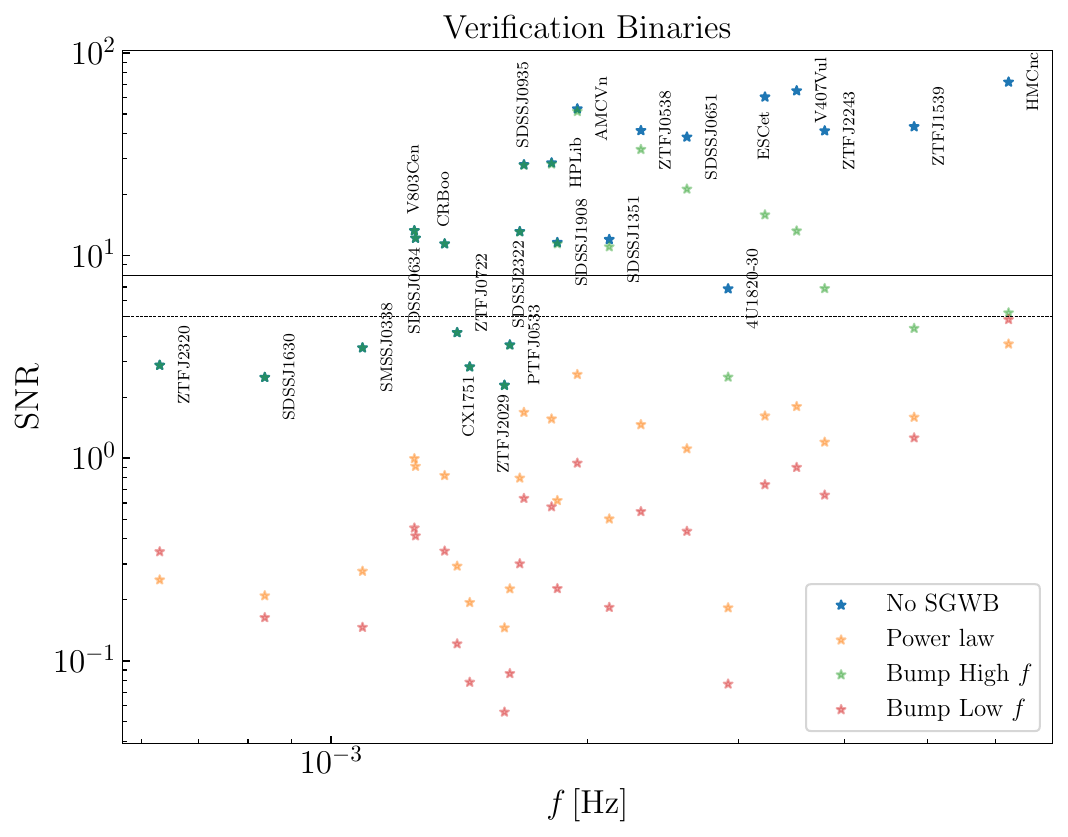}
	\caption{[Left] SNR histogram for Galactic Binaries from the LDC catalogue, for the different SGWB cases. We start from a reduced catalogue with purely instrumental $\mathrm{SNR}>5$, plotted in black. [Center] Percentage loss of detached and interacting binaries with ${\rm SNR}>5$ (with respect to the No SGWB case) as a function of frequency. [Right] Degradation of the SNR of the verification binaries. We only plot those that have ${\rm SNR}>5$ in the No SGWB case. In the left and right panels, dashed and solid lines indicate SNR $=5$ and $8$ respectively.}
	\label{fig:GBs_bgs}
\end{figure*}

As an illustrative case, we consider an event with characteristics similar to GW150914~\cite{LIGOScientific:2016aoc} and discuss the SGWB's impact on its detectability prospects with LISA. In practice, we choose the masses, $d_L$, and spins to match the measured values, pick the initial frequency, say $f_{\rm i}$, such that the signal exits the LISA band after 4 years of data taking, and randomize all the angular variables, i.e., inclination, sky position, and phase. The results of this analysis are shown in Fig.~\ref{fig:SOBBH_bgs}. Notice that the peak at low values in the SNR distribution, which appears in all cases, is mostly due to the dependency on $\iota$. Indeed, the SNR is minimal for $\iota = \pi / 2$, which corresponds to the peak of the distribution for the number of events.

Since with our choice for the initial frequency, the signal only appears in the high-frequency part of the LISA band (for reference, see Fig.~\ref{fig:bgs}), the SNR would largely be unaffected by the presence of a bump at low frequencies. For the same reason, the high-frequency bump here constitutes the most pessimistic case, which might reduce the SNR of a factor $\sim 3$. The PL case sits between the two other cases with a degradation in the SNR of a factor $\sim 2$. In both cases, the degradation is rather significant and might completely jeopardize the possibility of resolving individual SOBHBs with LISA.

In our analysis, we have assumed the initial frequency $f_{\rm i}$ for the signal such that the system leaves the LISA sensitivity band after 4 years of data taking. In principle, events could have different values of $f_{\rm i}$, which would impact the results of our analysis. Increasing $f_{\rm i}$, the system would spend less time in the LISA band, which reducing the SNR, but without modifying the qualitative behaviors for the different SGWB benchmarks. On the other hand, reducing $f_{\rm i}$ implies that the system does not leave the LISA frequency band after 4 years of data taking, leading to more interesting nontrivial considerations. In this case, the system only spans a range of frequencies between $f_{\rm i}$ and a maximal frequency\footnote{Notice that $f_{\rm e}$ decreases quite rapidly with $f_{\rm i}$. In particular, systems with very low $f_{\rm i}$ are still deep in the inspiral regime. Similar to GBs, these systems do not evolve much in frequency during the mission lifetime and can be effectively treated as monochromatic signals.}, say $f_{\rm e}$. In this case, depending on the specific choice of $f_{\rm i}$, the bump at low frequency (or at some intermediate position) might impact the SNR more than the High $f$ bump case. Similarly, assuming we stick to the choice of determining $f_{\rm i}$ so that the signal leaves the LISA band after 4 years of data taking, increasing the mass of the system would lower the merger frequency and speed up the frequency evolution, shifting  $f_{\rm i}$ to a lower frequency.  Even in this case, we expect some qualitative modifications of the conclusions drawn in this section.

\subsection{Galactic Binaries}
\label{sec:gal}
As a last example, we discuss the impact of large SGWBs on GB detections. For this purpose, we use the LISA Data Challenge (LDC) catalog for GBs based on the populations in~\cite{Korol:2017qcx,Korol:2020lpq, Nelemans:2013yg}. As for SOBHBs,
 we restrict our analysis to SNR considerations. Since the SNR of a specific source can only decrease in the presence of an SGWB, we start by computing the SNR for all the events in the catalog in the case where only the instrumental noise is present (i.e., we neglect the confusion noise) and drop all the events with SNR $< 5$, being undetectable even in the most optimistic scenario. We then compute the SNR for the reduced catalog for the ``No SGWB'' configuration and the three benchmarks discussed throughout this paper. The results of the analysis are shown in  the left panel of Fig.~\ref{fig:GBs_bgs}, where we show variations of the SNR distribution. Since most GB signals sit on the low end of the LISA frequency band (see Fig.~\ref{fig:bgs}), a bump at low frequency has the greatest impact on the SNR distribution of the sources, whereas a bump at high frequency has the least. Nevertheless, even in the latter case nearly 90\% of the GBs are lost at frequencies in the range $\sim[3,\,15]$ mHz. The PL case, while not as severe as the low-frequency bump, would still significantly reduce the number of sources LISA can resolve, decreasing from tens of thousands to hundreds. As for SOBHBs, if the degradation due to a large SGWB is present, it could dramatically threaten the prospects of resolving individual GBs with LISA.  These arguments are also made clear by looking at the central panel in Fig.~\ref{fig:GBs_bgs}, where we show the percentage loss of GBs with SNR$>5$ as a function of frequency, taking as a reference the No SGWB case. The plot shows that in the case of the Power law and the Bump low $f$ cases, we lose 100\% of the binaries for $f<6$ mHz. On the other hand, in the Bump high $f$ case, we never lose 100\% of the binaries in a frequency bin, and reach a maximum of $80$-$90$\% of lost binaries in the frequency range $f\in[0.003,\,0.015] $ Hz, with almost all the binaries outside this frequency range that remain detectable.

One major impact of this degradation would be on the detection of verification binaries~\cite{Stroeer:2006rx} plotted in Fig.~\ref{fig:bgs}. These are ultra-compact GBs with orbital periods shorter than a few hours, usually composed of degenerate stellar remnants such as white dwarfs, neutron stars, or SOBHs. Having already been detected electromagnetically and expected to be individually detected due to their strong GW signals, they are considered guaranteed sources for LISA that are crucial for facilitating the functional tests of the instrument and maximizing LISA's scientific output. However, only a few of them would be detectable in the presence of a significant SGWB with substantial support at frequencies $\lesssim 0.01$ Hz. For the 3 benchmark SGWBs considered in this work, we show explicitly the lost verification binaries in the right panel of Fig.~\ref{fig:GBs_bgs}.

As a final comment, we stress that the strong reduction in the SNR of each source would complicate the subtraction of signals from GBs from the data. As a consequence, we expect a (possibly) significant impact on the overall confusion noise due to these sources, which peaks at around $1$ mHz, see the small bump in the blue sensitivity curve in Fig.~\ref{fig:bgs}. Qualitatively, due to a less efficient removal of individual sources, we expect the signal to extend to higher frequencies and to have a larger amplitude. An updated estimate of the confusion noise would require techniques similar to the ones employed in~\cite{Karnesis:2021tsh}, which lay beyond the scopes of the current work.

\section{Discussion and conclusions}
\label{sec:conclusions}

Gravitational wave (GW) detectors are sensitive to two types of signals: deterministic sources, such as coalescing compact objects, and stochastic gravitational wave backgrounds (SGWBs), produced by the incoherent superposition of unresolvable sources or early Universe mechanisms. The latter appear as additional noise in GW detectors, potentially dominating the instrumental noise at certain frequencies and making deterministic sources harder to resolve. Well-motivated models of the early Universe predict an SGWB that could be significantly larger than LISA's instrumental noise level. The only existing constraints on the SGWB amplitude in the LISA frequency band are experimental, coming from Cosmic Microwave Background (CMB) and Big Bang Nucleosynthesis (BBN).

Our paper is based on the observation that current bounds on the SGWB still allow for amplitudes large enough to threaten the detection of deterministic sources. We conducted a comprehensive study on how a strong SGWB, saturating the CMB+BBN bounds, could degrade LISA's capability to detect deterministic sources.  Our results should be considered as the most pessimistic scenario for the LISA mission. Specifically, we analyzed the impact on three major astrophysical targets of LISA: Massive Black Hole Binaries (MBHBs), Stellar Origin Black Hole Binaries (SOBHBs), and Galactic Binaries (GBs). We modeled the SGWB energy density, $h^2\Omega_{\rm GW}(f)$, as either a narrow bump peaking at low frequency ($2.2\times 10^{-3}$ Hz) or high frequency ($2.2\times 10^{-2}$ Hz), or a flat power law across LISA's sensitivity band. While this approach is phenomenological, it reflects realistic scenarios: peaked SGWBs can be produced by various inflationary mechanisms~\cite{Braglia:2024kpo} or First Order Phase Transitions~\cite{Caprini:2024hue}, and nearly flat SGWBs are typically generated by cosmic strings~\cite{Blanco-Pillado:2024aca}, which are expected to fall in the LISA band if the recent PTA detection of a SGWB is interpreted in this context~\cite{Ellis:2023tsl,Lazarides:2023ksx,Figueroa:2023zhu,Ellis:2023oxs}.

Our results are summarized as follows.
\begin{itemize}
    \item {\bf MBHBs.} We conducted a comprehensive study on how the distribution of the SNR and the errors on the binary parameters, estimated via a Fisher analysis, are impacted by an SGWB. We examined three representative types of MBHB populations, each exemplifying different features. We further complemented this with a detailed Bayesian analysis of a pair of representative events, which we identified among the three populations. Our main findings are that a strong SGWB will reduce the number of detectable events, predominantly affecting populations with relatively small masses ($M_z/M_\odot\sim10^4-10^5$). The exact number of non-detectable events depends on the spectral shape of the SGWB. We also find that the errors on many binary parameters can increase by up to an order of magnitude. Interestingly, a large SGWB can strongly impact the sky localization, as we have observed through the appearance of bimodal distributions in the event longitude, even for relatively close events that would be well-localized without the SGWB.    
    \item {\bf SOBHBs.} For these sources, we only performed an SNR analysis, which was sufficient for our purposes. These types of signals are only affected by SGWBs with support at high frequencies. Given that these sources are already relatively faint, we found that the effect of the background is either to entirely prevent their detection or to have nearly no effect on them.    
       \item {\bf Galactic binaries.} For this type of source, we found results similar to those for SOBHBs. In this case, the effect of the SGWB is also of the "on-off" kind. However, as opposed to SOBHBs, these sources emit at lower frequencies, and SGWBs only marginally affecting the detections are those peaking at high frequencies.
\end{itemize}

In short, our results show that there would be a strong impact on LISA's scientific objectives in the pessimistic scenario (for detecting individual GW signals with LISA) where a primordial mechanism produces a loud SGWB. We find that the study of MBHBs would be significantly affected, although still possible to a more limited extent. On the other hand, SOBHBs and GBs could be rendered completely undetectable, potentially nullifying two of LISA's major objectives. While a detailed study of the dependence of our assessments on the properties of the SGWB is beyond the scope of this work, we point out that these considerations are often not taken into account when assessing the mission's capabilities. For example, it would be interesting to quantify,  the maximal SGWB amplitude leading to the minimal degradation in detecting the sources mentioned above. Beyond the sources we have studied, these important results warrant further investigation into the impact of the SGWB on the detection and characterization of other types of signals, such as Extreme Mass Ratio Inspirals or other burst-like signals.

To conclude, let us discuss the implications of our study on some of the major astrophysical and cosmological tests that the community hopes to perform by detecting the sources we studied. Beginning with MBHBs, the reduction in the number of detections and the degradation of the forecast errors will impact the astrophysical and cosmological tests which require a large statistical sample of events. These studies include population analyses, which provide insights into the formation and evolution of MBHs, and cosmological studies aimed at testing the expansion rate of the Universe and potential modifications to gravity. However, we expect that tests of the nature of BHs in the strong gravity regime or those relying on detecting electromagnetic counterparts could still be performed by detecting some very strong sources, which would be less affected by the SGWB.

Moving on to SOBHBs, a loud SGWB may completely ruin the prospects of detecting this class of sources with LISA. Thus, the most direct implication would be the loss of multiband GW detection~\cite{Sesana:2016ljz}. A secondary yet significant consequence would be the partial or even total masking of the SGWB generated by these objects. As demonstrated in~\cite{Babak:2023lro}, measuring such a signal could provide substantial insights into the behavior of these objects at high redshift, revealing properties of the population that complement those tested by LVK. A large SGWB would make it challenging to achieve this intriguing prospect.

Finally, our results on GBs suggest that a significant SGWB could potentially disrupt the program of verification binaries, which are expected to play a crucial role in the science verification of LISA data.

The findings of our paper highlight the crucial need to consider the potential impact of an SGWB on the detection of deterministic sources. Given the vast range of theoretical predictions, uncertainties related to the SGWB might outweigh those in the characterization of LISA's noise. We hope this study spurs further research to devise strategies that minimize the impact of an SGWB and fully exploit LISA's capabilities. On an optimistic note, even if the scientific goals related to the detection of deterministic sources would be somewhat compromised, detecting an SGWB of the amplitude considered in this paper would be groundbreaking. It would confirm its primordial origin and mark a significant milestone in physics, showcasing the remarkable potential of the LISA instrument.

\vspace{0.5cm}
\noindent

\begin{acknowledgments}
We thank Nicola Tamanini for interesting discussions in the early stages of this project. We also thank Valeriya Korol and Luca Reali for very useful comments on a draft of this paper. MP acknowledges the hospitality of Imperial College London, which provided office space during some parts of this project. MP also thanks the Center for Cosmology and Particle Physics (CCPP) at New York University for its hospitality during the initial and final parts of the project. 
\end{acknowledgments}

\end{document}